\documentclass[journal=jacsat,manuscript=article]{achemso}
\mciteErrorOnUnknownfalse
\usepackage[colorlinks=true,citecolor=blue,linkcolor=magenta]{hyperref}
\usepackage{graphicx}
\usepackage{epsfig}
\usepackage{color}
\usepackage[loose]{units}
\usepackage{soul}
\usepackage{hyperref}
\usepackage{cleveref}
\usepackage[caption=false]{subfig}
\usepackage[english]{babel}
\usepackage{letltxmacro}
\usepackage[binary-units=true]{siunitx}
\LetLtxMacro{\ORIGselectlanguage}{\selectlanguage}
\makeatletter
\DeclareRobustCommand{\selectlanguage}[1]{%
    \@ifundefined{alias@\string#1}
      {\ORIGselectlanguage{#1}}
      {\begingroup\edef\x{\endgroup
         \noexpand\ORIGselectlanguage{\@nameuse{alias@#1}}}\x}%
}

\definecolor{linkColor}{rgb}{1,0,0}
\hypersetup{pdfborder={0 0 0},colorlinks=true,urlcolor=linkColor,citecolor=linkColor}

\newcommand{\UCBMaterials}{Department of Materials Science and Engineering, University of California Berkeley, Berkeley, CA 94720}

\newcommand{\LBL}{The National Center for Electron Microscopy, Molecular Foundry, 1 Cyclotron Road, Berkeley, California 94720 USA}

\newcommand{\equal}{these authors contributed equally to this work}

\title{Simultaneous Successive Twinning Captured by Atomic Electron Tomography}

\author{Philipp M. Pelz}
\affiliation{\UCBMaterials}
\alsoaffiliation{\LBL}
\alsoaffiliation{\equal}
\author{Catherine K. Groschner}
\affiliation{\UCBMaterials}
\alsoaffiliation{\equal}
\author{Alexandra Bruefach}
\affiliation{\UCBMaterials}
\author{Adam Satariano}
\affiliation{\UCBMaterials}
\author{Colin Ophus}
\affiliation{\LBL}
\author{M.C. Scott}
\email{mary.scott@berkeley.edu}
\affiliation{\UCBMaterials}
\alsoaffiliation{\LBL}

\begin{document}
\maketitle
\date{\today}
%%%%%%%%%%%%%%%%%%%%%%%%%%%%%%%%% Abstract%%%%%%%%%%%%%%%%%%%%%%%%%%%%%%%%%%%%%%%%%%%%%%%

\textbf{
Shape-controlled synthesis of multiply twinned nanostructures is heavily emphasized in nanoscience, in large part due to the desire to control the size, shape, and terminating facets of metal nanoparticles for applications in catalysis. Direct control of the size and shape of solution-grown nanoparticles relies on an understanding of how synthetic parameters alter nanoparticle structures during synthesis. However, while outcome populations can be effectively studied with standard electron microscopy methods, transient structures that appear during some synthetic routes are difficult to study using conventional high resolution imaging methods due to the high complexity of the 3D nanostructures. Here, we have studied the prevalence of transient structures during growth of multiply twinned particles and employed atomic electron tomography to reveal the atomic-scale three dimensional structure of a Pd nanoparticle undergoing a shape transition. By identifying over 20,000 atoms within the structure and classifying them according to their local crystallographic environment, we observe a multiply-twinned structure consistent with a simultaneous successive twinning from a decahedral to icosahedral structure. We also observe a high degree of structural disorder in the nanoparticle, and a disordered crystal structure on the particle surface. Our results shed light on the transition mechanism for formation of icosahedral nanoparticle structures.}
\\

Multiply twinned particles (MTPs) are ubiquitous in solution-grown nanoparticle populations of face-centered cubic (fcc) metals. The ability to improve catalytic activity by controlling the exposed surface facet and strain states of the MTPs is a major driving force to understand their evolution during synthesis \cite{Choi_2015,Li_Cui_Ross_Kim_Sun_Yang_2017,Li_Cui_Ross_Kim_Sun_Yang_2017,Xia_Xiong_Lim_Skrabalak_2009,Narayanan_Cheng_Zeng_Zhu_Zhu_2015,Li_Qiang_Vuki_Xu_Chen_2011,Pietrobon_Kitaev_2008,Wang_2015}. The primary MTPs that appear in fcc metal nanoparticle populations are decahedra and icosahedra, which have an idealized structure described by assemblies of 5 or 20 tetrahedral subunits respectively, with subunits joined by twin boundaries of close-packed $\left<\mathrm{111}\right>$-terminated surface facets. However, other multiply twinned structures are commonly observed, either as transient structures during synthesis \cite{langille2011plasmon, Ma_Lin_Chen_Jin_2020} or final reaction products \cite{hofmeister1984habit,Hofmeister_2009}. 

Studies of MTP synthesis have identified many possible growth pathways to generate MTPs \cite{Xia_Xiong_Lim_Skrabalak_2009,Ma_Lin_Chen_Jin_2020, langille2011plasmon, langille2012stepwise}. Many efforts to create size and shape controlled MTPs center around controlling the population of nanometer-sized crystal seeds, which can uniformly grow into larger particles with the same shape \cite{Lim2007}. However, other structural evolution pathways are known to occur during colloidal growth of MTPs, such as successive twinning and oriented attachment \cite{Ma_Lin_Chen_Jin_2020, Song_Zhou_Lu_Lee_Nakouzi_Wang_Li_2020}. The  successive twinning growth process is unique to MTPs and refers to the additive growth of new tetrahedra to multiply twinned structures. This process can evolve single tetrahedra into a more complex MTP by island-to-tetrahedron growth on one of the facets of the single tetrahedra. 

The details of the successive twinning process are not clear. The decahedron to icosahedron successive twinning growth pathway is particularly difficult to quantitatively characterize using two dimensional imaging methods given the complex overlapping crystal grain structure of these MTPs. Some studies have indicated that icosahedra can evolve from fully formed decahedra \cite{langille2011plasmon,langille2012stepwise}, while others claim icosahedra generated through successive twinning require a partially formed decahedra \cite{tsuji2010stepwise}. Furthermore, the role of local defects and surface structures play during successive twinning have not been identified. 

\begin{figure*}[ht!]
    \centering
    \includegraphics[width=\textwidth,scale=0.5]{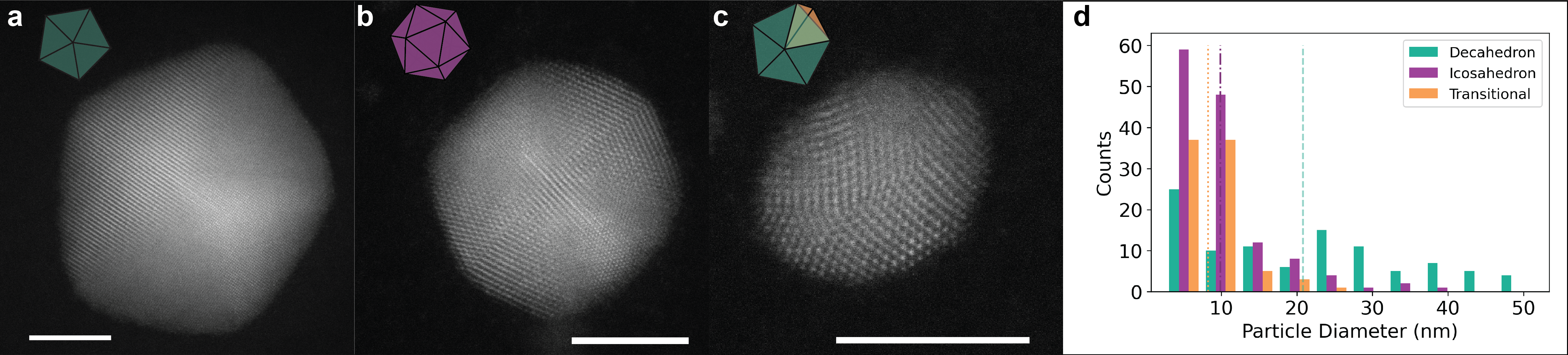}
    \caption{Representative images of the structures observed and histogram from the aliquot study. a) Sample decahedron. b) Sample icosahedron. c) Sample particle transitioning between decahedron and icosahedron. d) Histogram of the populations observed for the different structures as a function of size. Scale bar represents \SI{5}{\nano\meter}}.
    \label{fig:aliquot}
 \end{figure*}
 
Electron microscopy characterization has been critical in our understanding of MTP particle stability and growth. Several examples of successive twinning have been observed using electron microscopy in both static and liquid cell experiments \cite{langille2011plasmon,langille2012stepwise, Ma_Lin_Chen_Jin_2020}. However, most previous experimental electron microscopy studies have been limited to two dimensional imaging, such that many of the multi-tetrahedron stuctures present during successive twinning appear ambiguous. Atomic electron tomography (AET) is a method that utilizes high-resolution scanning transmission electron microscopy (HR-STEM) datasets to reconstruct the atomic-scale 3D structure of materials. Previous AET studies have resolved the structure, local defects and strain in icosahedral and decahedral metal nanoparticles \cite{Scott_Chen_Mecklenburg_Zhu_Xu_Ercius_Dahmen_Regan_Miao_2012,Goris_2015}, making it an ideal tool to resolve the complicated crystal structure of nanoparticles undergoing successive twinning.

In this work, we applied conventional HR-STEM to a population of Pd nanoparticles to determine the frequency of appearance of decahedra, icosahedra, and multiply twinned particles undergoing successive twinning, which we will refer to as multi-tetrahedron particles. AET was then used  to measure the atomic-scale structure of a representative particle undergoing successive twinning. AET revealed significant structural disorder within the particle.  The more than 20,000 atomic coordinates in 3D provided by AET analysis were further classified according to their local fcc or hexagonally close-packed (hcp) environment. This classification enabled atomic-scale 3D visualization of a simultaneous successive twinning process, where the particle was midway through a transformation from decahedron to icosahedron. We also observe stacking faults and other defects within the crystal grains of the particle, and a region of crystalline disorder on the surface of the particle.

\section{Results and Discussion}

\subsection{Population Statistics of Pd Nanoparticles}
To understand the relationship between MTP size and  structure, we first analyzed Pd nanoparticle populations using HR-STEM. Following previous work \cite{Lim2007}, we used an aqueous synthesis known to produce decahedral particles that employs polyvinylpyrrolidone (PVP) as a stabilizing agent and citrate as a reducing agent and capping agent. To broadly capture size and shape statistics, incubation time was varied from 1 hour to 24 hours (Methods).  Population statistics of the synthesized nanoparticles were manually determined from 691 particles identified in the HR-STEM data. Figure \ref{fig:aliquot} summarizes the results of the HR-STEM study. The results confirm the presence of decahedra, iscoahedra, and multi-tetrahedron particles that appear to be undergoing a successive twinning process. Figures \ref{fig:aliquot}a-c show sample micrographs of these structures. The size and shape distributions presented in Figure \ref{fig:aliquot}d show a clear size trend between the three structures. The icosahedra and successive twinning structures have size distributions skewed towards smaller sizes, with average sizes of \SIrange[range-phrase=\ $\pm$\ ]{9.87}{0.57}{\nano\meter} and \SIrange[range-phrase=\ $\pm$\ ]{8.28}{0.41}{\nano\meter}, respectively. The decahedra show a more uniform distribution with an average size of \SIrange[range-phrase=\ $\pm$\ ]{20.8}{1.39}{\nano\meter} . The synthesis conditions used for this study were a relatively high concentration of reducing agent as well as a stabilizing agent (PVP), and should result primarily in decahedra \cite{Xiong2007}. PVP, however, primarily interacts with nanoparticles larger than \SI{10}{\nano\meter} \cite{Koczkur2015}. If the PVP interaction is critical to stabilizing growing decahedra this may explain why multi-tetrahedron and icosahedron particles are primarily observed at sizes below approximately \SI{10}{\nano\meter}. 
\begin{figure}[hbt!]
    \centering
    \includegraphics[width=3.4 in]{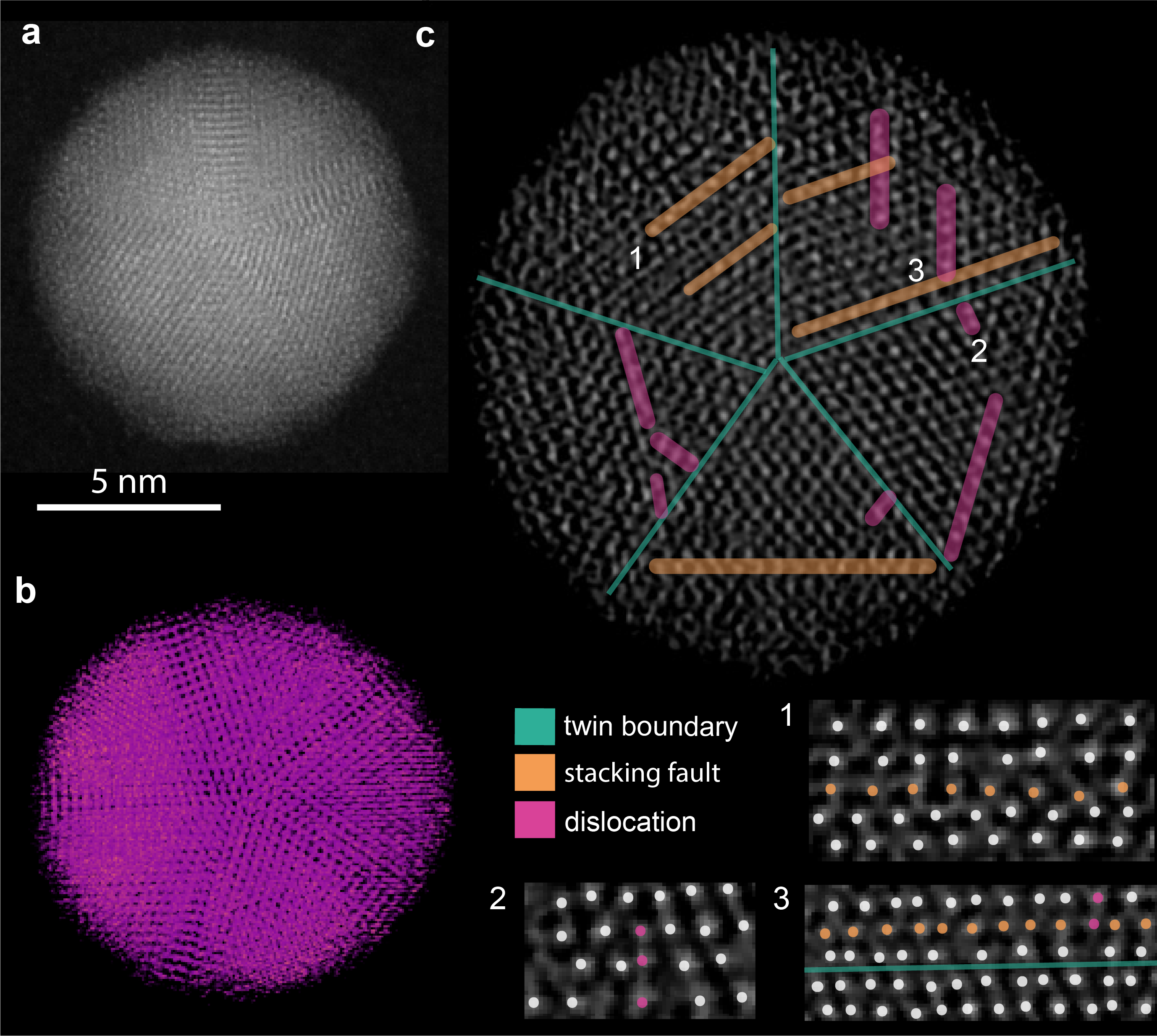}
    \caption{ a) Sample STEM micrograph from the tilt series. b) Volumetric rendering of the tomographic reconstruction. c) An example slice through the atomic volume normal to the five-fold axis with various defects highlighted.}
    \label{fig:defects}
 \end{figure}
 
 \subsection{Atomic Electron Tomography of a Pd MTP}
To obtain a detailed understanding of the three-dimensional atomic structure of an MTP undergoing successive twinning, we performed AET on a representative Pd particle. We used aberration-corrected STEM to collect a tilt series of 49 images of an approximately 10 nm diameter particle, with tilt angles ranging from -61 to 64$^{\circ}$  (Supp.  Fig. 1). Images taken before and after the tilt series acquisition indicate that the particle's structure did not change during imaging (Supp.  Fig. 2). While the particle's shape and crystal structure were largely consistent with a decahedron (Fig. \ref{fig:defects}a), several images in the tilt series indicate the presence of additional crystalline grains in the nanoparticle, suggesting that the particle is undergoing successive twinning (Supp.  Fig. 1). After denoising and aligning the tilt series (Methods), we reconstructed the volume, shown in Figure \ref{fig:defects}b, using a FASTA based reconstruction algorithm (Methods). To assess the consistency between the final reconstructed volume and the input projections, we define the R-factor as the pixel-wise difference of the absolute values of the measured and calculated projections from the reconstructed volume, normalized by the intensities of the measured projections. The R-factor of the final reconstruction was 8.1\%, which is consistent with other reported AET reconstructions \cite{Yang_Zhou_2021}.

\begin{figure*}[ht!]
    \centering
    \includegraphics[width=\textwidth,scale=0.5]{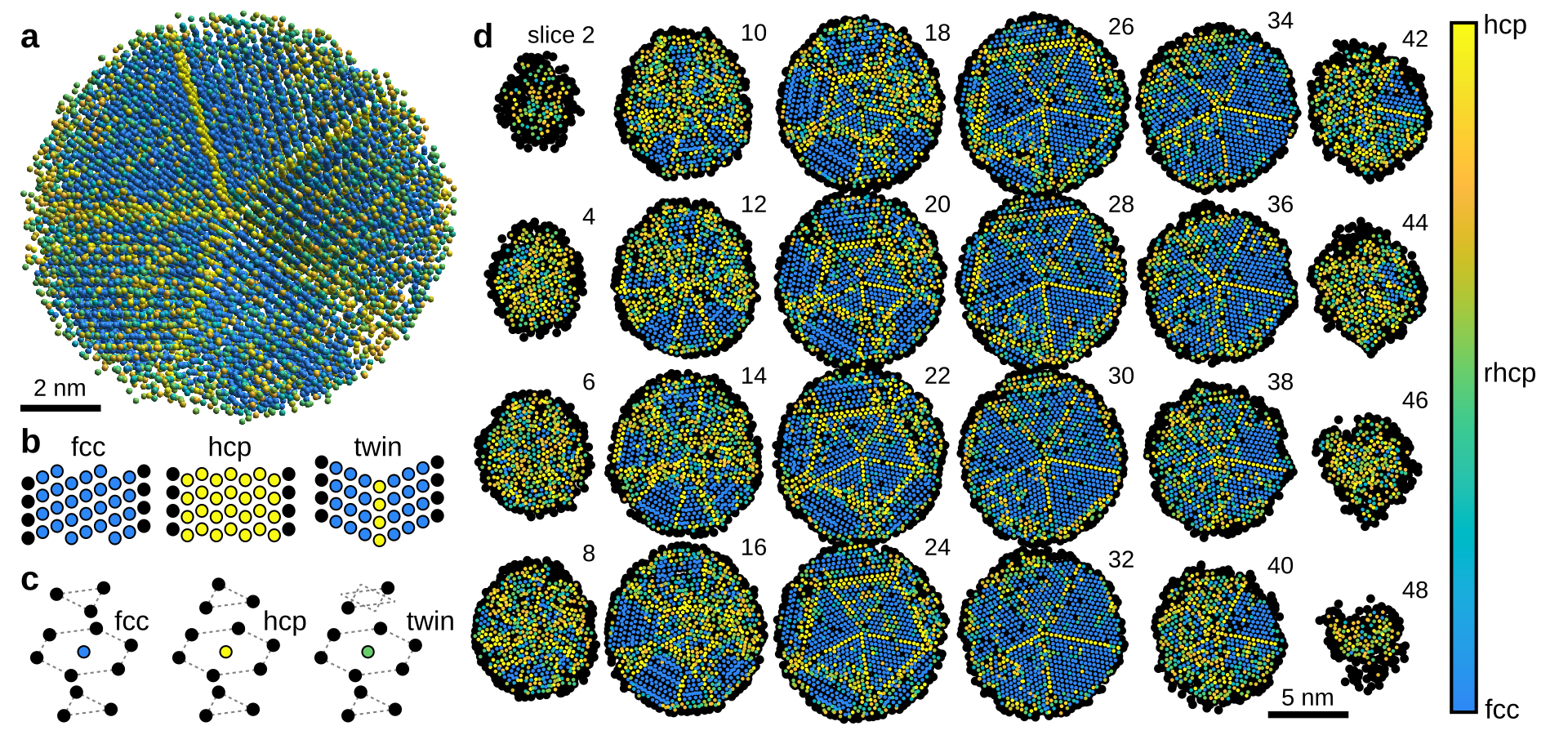}
    \caption{Crystallographic ordering of the atoms in the Pd MTP. (a) Traced atom positions with coordination numbers of 12, colored by fcc to hcp ordering.  (b) Atomic stacking arrangement of fcc, hcp, and a twinned structure view from the side. (c) 3D atomic neighborhood of sides with fcc, hcp, and rhcp ordering. (d) Slices throughout the MTP showing the crystallographic ordering for each atom with at least 10 neighboring sites.}
    \label{fig:coords}
 \end{figure*}

After the final reconstruction volume was obtained, atomic positions were determined using an iterative 3D Gaussian fitting procedure. Atomic locations were included or excluded based on unsupervised clustering of the atoms based on atom intensities and radial distribution function of the atoms. Subsequently we determined the atom set that maximizes the Fourier Ring Information \cite{van_Heel_Schatz_2020} between a linear image generated from the atom positions and the measured projections (see Methods for a detailed description). Using this method, we determined the atom positions of 20~632 atoms, shown in Figure \ref{fig:coords}a.

A qualitative analysis of the structures present in the reconstructed volume, shown in Figure \ref{fig:defects}b, reveals a large number of defects in the structure. To illustrate the large number of both stacking faults and dislocations in the particle a 0.25 $\textrm{\r{A}}$-thick slice through the volume with defects highlighted is presented in Figure \ref{fig:defects}. Many of the stacking faults and associated dislocations are observed adjacent to the twin boundaries, which was also observed in with molecular dynamics simulations of five-fold twinned metal nanowires \cite{Zhou_Fichthorn_2014}. However, defects and disorder are not restricted to regions around the twin boundaries and can be found within the tetrahedral subunit bulk, as shown in defect 3 in Figure \ref{fig:defects} c). 
The number of stacking faults and edge dislocations observed in the reconstructed particle is much higher than in previous studies \cite{Hofmeister_1991,Goris_2015}, and we also see many more defects than predicted for nanomaterials of this size \cite{Zhou_Fichthorn_2014}. The stacking fault and twinning energy of fcc metals is in general low \cite{Rosengaard1993}, implying facile formation of stacking faults and twins during nanoparticle growth. Therefore, a possible explanation for the high number of defects in the particle is that the stacking fault mobility is slow compared to the rate of adatom addition to the particle. Another explanation for the high number of defects relies on the unique, and highly strained, structure of MTPs. Ideal fcc tetrahedra cannot be tiled into a decahedron or icosahedron in a way that is space filling \cite{Bagley_1965}. MTP decahedra and icosahedra, consequently, must contain a high degree of internal strain because the crystal lattice must accommodate the missing volume \cite{ino1969stability, Howie_Marks_1984}. Defects, such as stacking faults, are one mechanism for stress relief in MTPs \cite{Howie_Marks_1984}. Therefore, it is likely that the presence of the observed defects is also, at least in part, due to stress relief, and is consistent with  inhomogeneous strain within the particle.

\subsection{Crystal Structure of Successively Twinning Particle}
  
To better understand the crystal structure in the reconstructed particle we classified the traced atom coordinates according to their crystallographic coordination. First, all sites were sorted according to their coordination number using a cutoff radius of \SI{3.75}{\angstrom}. Atoms with coordination numbers below 10 were classified as surface sites, or highly disordered regions if they were fully contained in the nanoparticle. Most remaining sites in the particle bulk were primarily arranged into 4-atom tetrahedra. The most common structural packing of these tetrahedra were either fcc or hcp ordering. However, a large degree of disorder was present in many regions of the MTP on length scales larger than 4-atom clusters.

In order to determine the large scale structure of the MTP, we used a polyhedral matching algorithm inspired by Larsen et al.~\cite{larsen2016robust} to determine the degree of fcc and hcp ordering of each atomic shell. In this dataset, we found that polyhedral matching produced much more robust identification of the differently ordered MTP regions than other methods such as local bond order parameters \cite{steinhardt1983bond, lechner2008accurate} or common neighbor analysis \cite{faken1994systematic} (Supp. Fig. 5). A detailed description of the classification protocol is found in the Methods section. Briefly, for each valid atomic site we define polyhedra with vertices that correspond to fcc or hcp stacking geometries and fit them to the measured positions of the surrounding 12 atoms, depicted in Figure \ref{fig:coords}c. After rotating the polyhedra to minimize the distance between the ideal and measured atomic positions of the 12 nearest neighbors, we compute an order parameter for both fcc and hcp ordering (Methods). We then keep only the maximum order parameter for both fcc and hcp polyhedra. These values are used to determine the local ordering.

The results of this classification algorithm are shown in Figure \ref{fig:coords}. The 3D atomic coordinates with 12 nearest neighbors are shown in Figure~\ref{fig:coords}a, where each site is colored according to  the difference between the hcp and fcc order parameters defined above, from -1 to 1. The colors are strongly bimodal, indicating that the majority of sites possess either highly ordered hcp or fcc arrangements. The overall 5-fold symmetry of the nanoparticle is immediately obvious, though a significant amount of disorder is present both on the particle surface and in the bulk structure. 

To visualize the structural ordering of the entire structure, we have plotted every other atomic slice of the particle in Figure \ref{fig:coords}d, where the atoms are again colored by the difference between the hcp and fcc order parameters. Surface atoms are shown in black, as many of these sites do not have enough nearest neighbors to distinguish between fcc and hcp ordering. Figure \ref{fig:coords}b shows the two endpoint order parameters, the ideal fcc and hcp structures, viewed from the side. Figure \ref{fig:coords}b also shows a third structure motif which appears in many locations in Figure \ref{fig:coords}d, a twinned fcc structure, where a single line of atoms possessing high hcp ordering separates two fcc grains with mirrored structures. Figure \ref{fig:coords}c shows the 3D atomic arrangement of the two ideal polyhedra with fcc and hcp ordering, as well as a third structure which has the same order parameter for both structures (10 out of 12 possible sites agree with each class of polyhedron). This structure is labeled as random hexagonal closed packed (rhcp), following other studies which have used this label for packing falling between fcc and hcp \cite{harke2008three}.

\begin{figure*}[hbt!]
    \centering
    \includegraphics[width=0.8\textwidth,scale=0.5]{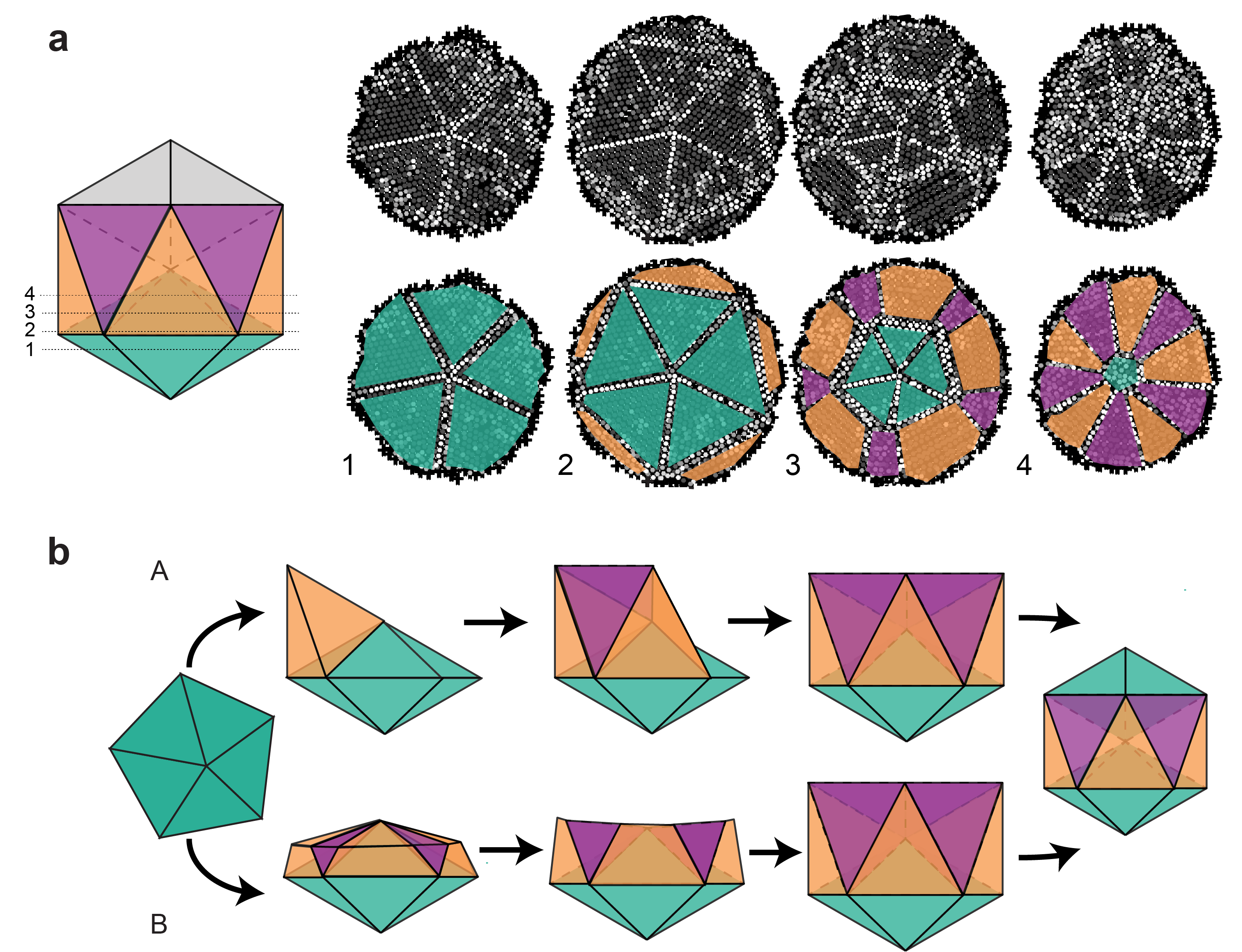}
    \caption{a) Sample slices from the labeled atomic coordinates shown in Figure \ref{fig:coords} labeled with how they would correspond to slices through an icosahedron. b) Schematic of the two pathways observed for the successive twinning process.}
    \label{fig:succesive_twinning}
 \end{figure*}

 The twinning structure obtained reveals that the reconstructed particle contains a core decahedron, with additional, partially formed tetrahedra on top. The resultant structure is consistent with a partially formed icosahedron (Figure \ref{fig:succesive_twinning}a.) Considering the arrangement of partially formed tetrahedral subunits atop the decahedral core of the reconstructed nanoparticle, shown in Figure \ref{fig:succesive_twinning}a, the particle seems to be captured in the midst of successively twinning, and has partially transformed into an icosahedron. There are ten additional partially formed tetrahedra, forming two rows above the core decahedron. The tetrahedra closer to the decahedron are more fully formed.  These nucleating tetrahedral units are highlighted in Figure \ref{fig:succesive_twinning}a.

Prior observations of successive twinning have been in particles on the 100 nm scale and have shown tetrahedron by tetrahedron growth \cite{hofmeister1984habit,langille2011plasmon,langille2012stepwise} illustrated schematically in Figure \ref{fig:succesive_twinning}b, pathway A. However, the successive twinning we observe, illustrated in Figure \ref{fig:succesive_twinning}b pathway B, is better described as a simultaneous process, where the tetrahedra comprising the center portion of the icosahedra grow at the same time. This process more closely resembles the successive twinning process predicted for metal nanoclusters, where   coordinated formation of hcp layer across a decahedral surface leads to simultaneous twinning  \cite{Baletto_2001_adatoms_HCP,Baletto_2005_review}. It is known that decahedra have more hcp adatom sites than fcc \cite{Baletto_2001_adatoms_HCP}, making hcp island growth, and therefore the growth of stacking faults and twin boundaries, more probable, especially if there is limited surface diffusion. Our observations also directly confirm addition of multiple tetrahedra to a decahedral particle as a route towards an icosahedral structure.
 
 \subsection{Disorder close to the surface}
 
 In addition to illustrating the twinning and defect structures present, our traced atomic coordinate classification protocol also reveals a large degree of disorder close to the bottom and top surface of the Pd nanoparticle. An ideal decahedral particle would exhibit $\left<\mathrm{111}\right>$ terminated surface facets. The top and bottom surface slices shown in the first and last column of Figure \ref{fig:coords}c show that the particle does not have $\left<\mathrm{111}\right>$ terminating facets. Instead, this surface of the particle contains a mixture of coordination types, including a significant fraction of random hexagonally close-packed (rhcp) coordinated atomic sites whose relation to fcc and hcp stacking is shown in Figure \ref{fig:coords}a. rhcp stacking has previously been observed during palladium nucleation and is further evidence of the non-equilibrium structure of these growing MTP particles \cite{Xia_Xiong_Lim_Skrabalak_2009}. The surface disorder is also an indication that surface diffusion on the particle is relatively slow \cite{Xia2013}.

\section{Conclusion}

In this study we have determined the 3D atomic positions of over 20,000 atoms in a multiply twinned palladium nanoparticle. We have found that the structure derived from the aqueous synthesis contains substantially more defects than would be expected from previous computational and experimental studies. We directly observe a simultaneous successive twinning process wherein a decahedral particle is transitioning directly to an icosahedron. Based on our HR-STEM studies, we suspect this process would only occur in small decahedral nanoparticles under our reaction conditions, as after a certain size the structural directing PVP will have a stronger influence on growth.

The complex structure observed has implications for nanoparticle functionality. The lack of $\left<\mathrm{111}\right>$ terminating facets will significantly alter catalytic reactivity in a Pd particle. Similarly, the high number of defects in the particle's interior structure will change the surface strain states of the particle, which have will also affect catalytic activity \cite{Choi_2015,Huang_2017}. Therefore, the combination of HR-STEM and AET used in this study provide unique insight into the structure and functionality of MTPs.

\newpage
\section{Methods}
\subsection{Sample preparation}
\label{subsec:sample_prep}
All reagents used in this synthesis were purchased from Sigma Aldrich. The Pd nanoparticle reaction was prepared based on the method reported by \citep{Lim2007} with slight modifications. We used the conditions reported to synthesize a population of primarily decahedral particles. Briefly, a \SI{15}{\milli\liter} three-necked flask was rinsed 3 times with MilliQ water and acetone, then dried. The flask was equipped with a reflux condenser and teflon-coated magnetic stir bar. A \SI{4}{\milli\liter} aqueous solution containing  poly(vinyl pyrrolidone) (PVP, 55,000 MW) and citric acid was transferred to the flask and heated to 90$^{\circ}$C while stirring using a heating mantle. Upon reaching 90$^{\circ}$C, A \SI{1.5}{\milli\liter} solution of Sodium tetrachloropalladate was rapidly added to the flask. For the aliqout study, samples were taken at 1, 3, 6, 8, 10, and 24 hours. The final product was isolated by adding a \SI{300}{\micro\liter} aliqout of acetone to \SI{100}{\micro\liter} nanoparticle solution in a clean microcentrifuge tube and centrifuged for 30 minutes at 13,000 rpm. The supernatant was decanted and the pellet was re-suspended in \SI{300}{\micro\liter} ethanol. The wash and rinse was repeated 2 times, and the purified particles were stored in MilliQ water.
Aqueous solution of Pd particles was deposited via nebulizer on a SiN window. The tomography study utilized a sample taken after incubating for 8 hours.
\subsection{Data Acquisition}
\label{subsec:data_acquisition}
Several tomographic tilt series were acquired from Pd nanoparticles using the TEAM 0.5 microscope and TEAM stage \cite{Ercius_Boese_Duden_Dahmen_2012} at the National Center for Electron Microscopy in the Molecular Foundry. Images were acquired at 200 kV in ADF-STEM mode with a \SI{25}{\milli\radian} convergence semi-angle (resulting in a probe size of \SI{0.8}{\angstrom}), \SI{41}{\milli\radian} and \SI{210}{\milli\radian} detector inner and outer semi-angles, and a beam current of \SI{4}{\pico\ampere}. The tilt series was collected at 49 angles with a tilt range of \num{64} to \num{-61} degrees. Two images per tilt angle were measured with \SI{3}{\micro\second} dwell time with a scan rotation of 0 and 90 degrees to minimize image blurring. Owing to imperfections in the calibration of the x- and y- scanning coils in the microscope’s STEM scanning system, an additional correction was applied to the images to ensure square pixels. This scan distortion was measured using a standard sample under the same imaging conditions, and corrected using the method described by Ophus et al.~\cite{Ophus_Ciston_Nelson_2016}.

\subsection{Image denoising}
\label{subsec:denoising}
The drift-corrected images were denoised with the BM3D algorithm \cite{Danielyan_Katkovnik_Egiazarian_2012}, with Anscombe variance-stabilizing transform and its inverse applied before and after denoising \cite{Yang_Chen_Scott_Ophus_Xu_Pryor_Wu_Sun_Theis_Zhou_2017}.

\subsection{Tomographic reconstruction}
\label{subsec:tomographic_recons}
After denoising, the 49 images were projected onto the tilt axis, and aligned with sub-pixel cross-correlation among the 1D-curves. A constant intensity scaling factor was fitted to the 1D curves to account for nonlinearities in the intensity at high tilt angles. Then the images were aligned to neighboring tilts with 2D subpixel cross-correlation. From the initially-aligned tilt series a 3D reconstruction was performed using the fast adaptive shrinkage-thresholding algorithm (FASTA), an accelerated gradient algorithm with adaptive stepsize for faster convergence \cite{Goldstein_Studer_Baraniuk_2014}. To compute the forward- and backward projections, we used the generalized ray transform interface of the Operator Discretization Library \cite{ODL_2018} in a 3D parallel-beam Euler geometry with an GPU-accelerated backend of the ASTRA tomography toolbox \cite{ASTRA_2015}. To increase the accuracy of the projections, we used the trilinear interpolation feature of the ASTRA library to compute the forward and inverse Ray-transforms. The code is freely available at \href{https://github.com/PhilippPelz/fasta-tomography}{this url}. To minimize the translational and angular misalignments, we use a projection matching approach \cite{Dengler_1989} with simulated annealing, where all three Euler angles are varied by a randomly picked value in the range of \num{-0.5} to \num{+0.5} degrees and the calculated projection error compared with the current projection error after a full reconstruction. The lowest-error angles are then used as new initial angles for the next tomographic reconstruction. This process is repeated for \num{10} outer iterations and the random Euler angle perturbation reduced linearly every iteration. Using this approach, the reconstruction converged to an R-factor of \SI{8.1}{\percent}

\subsection{Atom tracing}
\label{subsec:atom_tracing}
The 3D atomic positions of the Pd atoms were determined using the following procedure based on the code provided by Ren et al.~\cite{ren2020multiple}. (I) all local intensity maxima were identified from the 3D reconstruction and added to a candidate list. From the initial candidate list, peaks which were within a minimum distance of \SI{2.2}{\angstrom} of a higher-intensity peak were deleted. (II) The initial list of peak positions was refined by fitting a 3D Gaussian function to each peak after subtracting neighboring peaks within a maximum radius of \SI{4}{\angstrom}. Using this initial atom candidate list, we added, refined, and merged new unidentified peaks for 4 iterations in the following order:
(III) Subtract the fitted Gaussians of all current peak candidates from the reconstruction volume. (IV) Add new candidate peaks over an intensity threshold of \num{50} to the candidate list. (V) Refine the positions of all atom candidates as in (II) for 4 iterations. (VI) Merge peaks that are closer than a minimum distance of \SI{2.2}{\angstrom} after position refinement. (VII) Refine the positions of all atom candidates as in (II) for 4 iterations. (VIII) go to (III) if iterations not done. (IX) A final set of 4 positions refinement iterations as in step (II) yielded the final set of coordinates of 22412 candidate atoms. Previous work then used an atom flipping procedure to eliminate low-intensity atom candidates whose addition does not decrease the experimental error. 

We found that some low-intensity atoms were in the center of the particle, therefore we included additional features to eliminate non-atoms from the atom candidate list. For each atom candidate we computed the radial distribution function (RDF) and split it up into 7 radial regions corresponding to sections between the peaks of the total RDF. The sectioned RDF together with the fitted peak intensity, sigma, and voxel intensities of the peak in a \SI{2}{\angstrom} radius then formed a X-dimensional feature vector for each atom candidate. We then used the Uniform Manifold Approximation and Projection (UMAP) \cite{McInnes_Healy_Melville_2020} method to project the feature vector onto a 2-dimensional manifold, shown in the Supplementary Information (Supp. Fig 3). On one end of this manifold we identified disordered surface atoms with low intensity. We then used a Bayesian Gaussian Mixture Model \cite{scikit-learn} to classify the candidate list in the 2-dimensional manifold into two classes of potential atoms and non-atoms. Then we ranked the candidates according to their probability to belong to the non-atom class. To select a threshold for including atoms in the non-atom class and we performed ADF-STEM simulations (see the following section) of our tilt series with different probability thresholds for including atoms in the non-atom class. The ADF-STEM simulations were then compared with the Fourier Ring Information (FRI) criterion \cite{van_Heel_Schatz_2020} to find the threshold that extracts the largest amount of information from the experimental data. The results are shown in Supplementary Information Figure 3. We found that a threshold of \SI{99.994}{\percent} extracted the maximum amount of FRI from the data. 

We then finalized the set of atoms by performing the following atom-flipping procedure to determine if an atom within the bracket from \SI{99.99}{\percent} to \SI{99.997}{\percent} should be added. For each atom that falls in this probability range we compute a tilt series with and without that atom and compare the resulting FRI, using a simple linear image formation model, with the image being a linear sum of 3D Gaussian distributions for each atom, and the standard deviation determined from the atom fitting procedure. If addition of an atom increases the FRI we include it in the list of atoms, otherwise we exclude it. This procedure yields a final set of 20~632 Pd atoms in the particle.

\subsection{STEM simulations and tracing precision}
% \label{subsec:atom_tracing}
For atom classification and to evaluate the precision of the atom tracing procedure, we re-created the tilt series of 49 projections with the refined experimental Euler angles from the traced coordinates with quantum mechanical STEM simulations using the PRISM algorithm\cite{Ophus_2017} implemented in the Prismatic simulation software  \cite{prismatic}. A total of 49 cubic super cells of size (\SI{11}{\nano\meter})$^3$ was created. The final atomic model was placed within the super cells. Individual super cells were divided into \SI{2}{\angstrom} slices along the beam direction and sampled with a pixel size of \SI{6.2}{\pico\meter} in the transverse direction. The experimental parameters of \SI{200}{\kilo\volt} high tension, \SI{25}{\milli\radian} convergence semi-angle, \SI{41}{\milli\radian} and \SI{210}{\milli\radian} detector inner and outer semi-angles, \SI{0}{\milli\meter} $C_3$ aberration were used for the simulation. We employed a Fourier interpolation factor of 10 in the PRISM algorithm and simulated 8 frozen phonon configurations. We matched the probe step to the reconstruction voxel size of \SI{25}{\pico\meter}. Each simulated ADF-STEM image was convolved with a Gaussian function to simulate incoherent source spread and other incoherent effects and minimize the difference to experimental images. A 3D volume was then reconstructed from the simulated tilt series with the FASTA algorithm described above and a new model was obtained by using the same atom tracing procedure. The new atomic coordinates were rotated and translated to minimize the global position deviations between the models and then atoms were matched between the models and a root mean square displacement calculated between the models, with a radial search cutoff of \SI{1.5}{\angstrom} around each atom. A histogram of the atomic deviation between the common pairs is shown in Supplementary Information Figure 4, indicating a mean deviation of \SI{32.2}{\pico\meter}. This is slightly higher than the deviation of previous AET studies \cite{Yang_Chen_Scott_Ophus_Xu_Pryor_Wu_Sun_Theis_Zhou_2017}, which we attribute to the relatively low number of projection measurements available relative to the size of the reconstructed particle.

\subsection{Atom Classification}
% \label{subsec:atom_classification}
To classify traced atoms according to their crystallographic coordination, we first generated polyhedra with 12 vertices $\mathbf{p}_j$ arranged in both fcc and hcp stacking geometries with nearest neighbor spacing equal to the mean measured value of \SI{2.93}{\angstrom}. These polyhedra are rotated to 1026 orientations roughly evenly spaced on 1/12th of the unit sphere. For each atomic site, these polyhedra were rotated using a matrix $\mathbf{m}$ to minimize the total distance between their (ideal) coordinates and the nearest relative atomic site using the iterative closest point (ICP) algorithm \cite{kjer2010evaluation}. Finally we compute an order parameter for each polyhedra at each site $s_k$ equal to
\begin{equation}
    s_k = \sum_{j=1}^{12} \rm{max}\left(
    1 - 
    \frac{| \mathbf{r}_{j} - \mathbf{r}_{k} - \mathbf{m} \, \mathbf{p}_j |}
    {d_{\rm{max}}}, 0
    \right),
\end{equation}
where $ \mathbf{r}_{j}$ is the position of the $j$'th neighboring coordinate to site $k$ at position $ \mathbf{r}_{k}$, and $d_{\rm{max}}$ is maximum allowed distance of a site from an ideal position, which we set equal to half the average nearest neighbor distance of \SI{1.47}{\angstrom}. This cost function can generate values of 0 to 12, where a value of 12 indicates perfect alignment between the polyhedral template.
We keep only the maximum order parameter (best agreement) for both fcc and hcp polyhedra, and use these values to determine the local ordering.\\
We have also implemented bond-order parameter analysis by spherical harmonics \cite{boo}. A comparison of the different crystalline order parameter calculation methods is shown in Supplementary Figure 5. Overall, we found the polyhedral matching analysis more robust to experimental noise.

\section*{Acknowledgments} 

We are grateful to P. Ercius for help with the calibration of the collection angles of the ADF detector. Work at the Molecular Foundry was supported by the Office of Science, Office of Basic Energy Sciences, of the U.S. Department of Energy under Contract No. DE-AC02-05CH11231. C.O. is supported by the USA Department of Energy Early Career Research Award program. P.M.P and M.C.S are supported by the Strobe STC research center, Grant No. DMR 1548924. C.K.G. is supported by the National Science Foundation Graduate Research Fellowship under Grant No. DGE-1752814.

\section*{Author contributions} 

M.C.S. conceived the overall project. A.B. and A.S. synthesized the nanoparticles. C.K.G. collected particle size data and tilt series. P.M.P. reconstructed the particle. P.M.P. and C.O. performed the atom tracing. P.M.P. performed STEM simulations. P.M.P. and C.O. performed precision calculations. C.O. implemented ordering classification. C.K.G. and M.C.S. wrote the manuscript. All authors commented on the manuscript.

\section*{Supplementary Information Available}
Supplementary information contains:
\begin{itemize}
    \item Full tilt series
    \item Before and after tomography micrographs of particle
    \item Plots for atom tracing
    \item STEM Simulations and Tracing Precision
    \item order parameter comparison
\end{itemize}

\nocite{*}
\bibliography{sample}
\section{Supplementary Information}
\subsection{Tilt Series}
\begin{figure*}[ht!]
    \includegraphics[width=\textwidth]{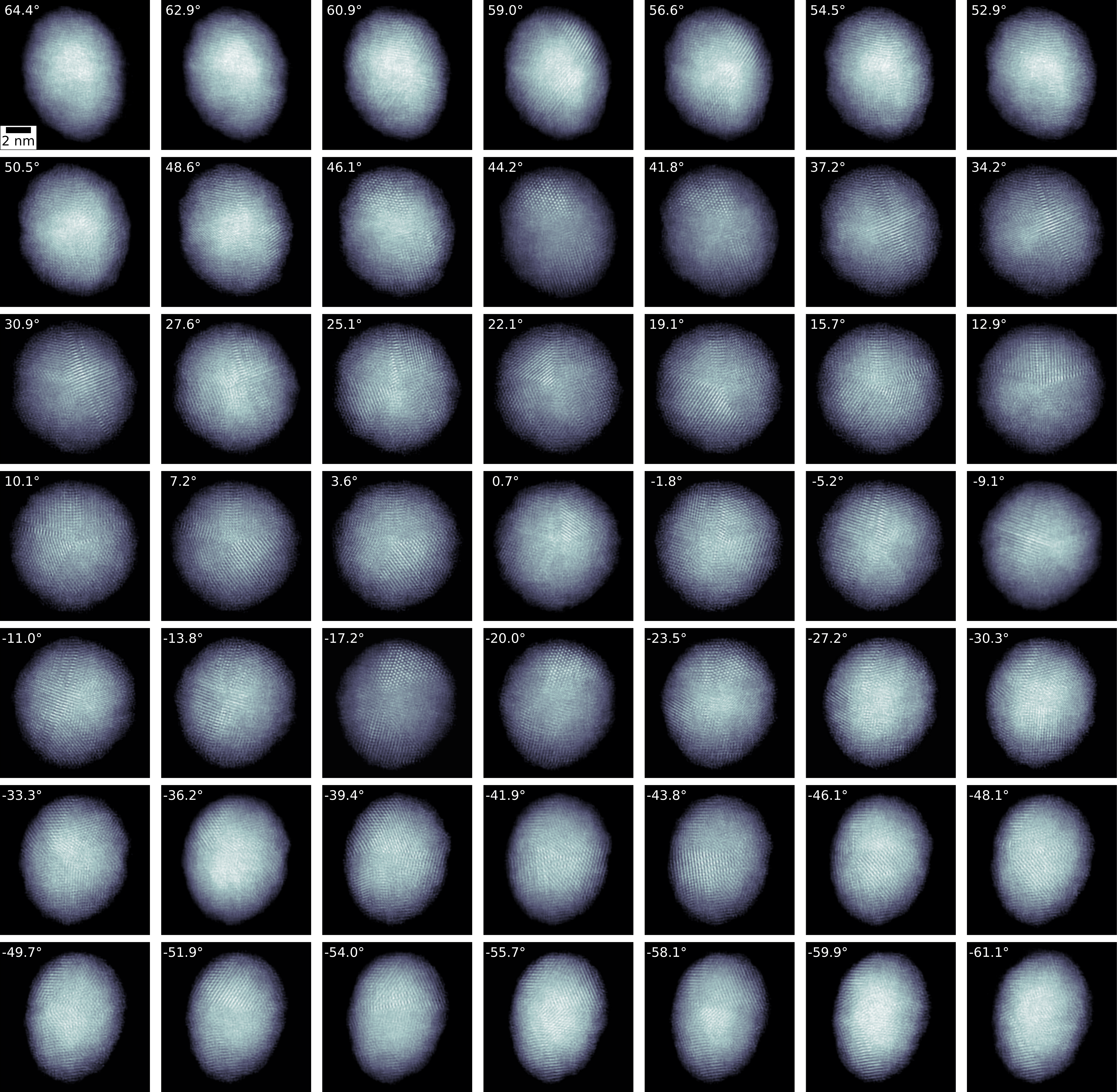}
    \caption{\label{fig:tilt_series} Tomographic tilt series of the multiply-twinned Pd nanoparticle. The 49 projection images with a tilt range from \num{64} to \num{-61} degrees (show at top left of each panel) were measured using ADF-STEM.}
\end{figure*}

\newpage
\subsection{Before and After Tomography}
\begin{figure*}[hbt!]
    \centering
    \includegraphics[width=0.6\textwidth,scale=0.5]{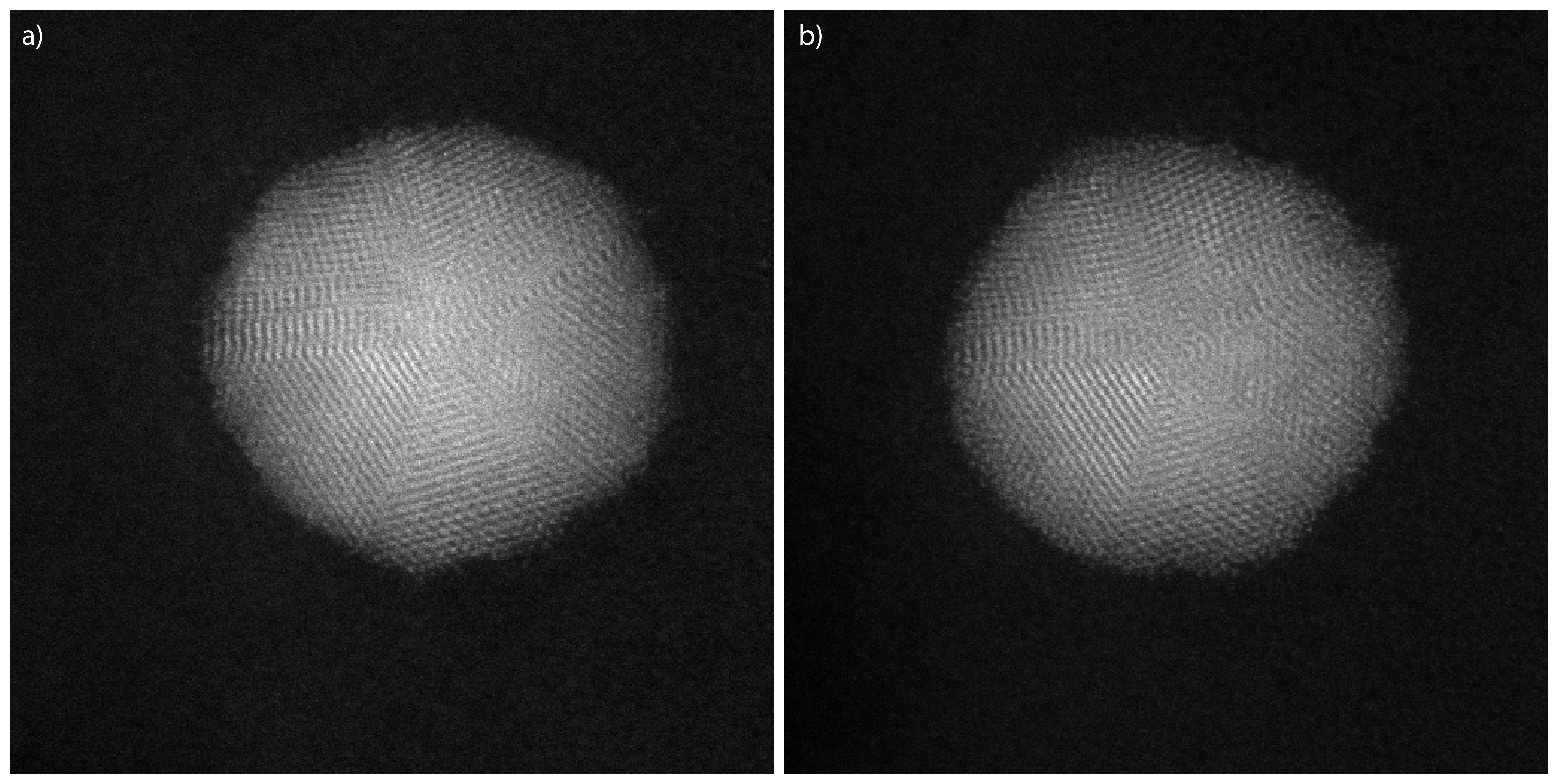}
    \caption{a) Particle before and (b) after tomography experiment.}
    \label{fig:before_and_after}
 \end{figure*}
 \newpage
 \subsection{Atom Tracing}
 \begin{figure*}[ht!]
    \includegraphics[width=\textwidth]{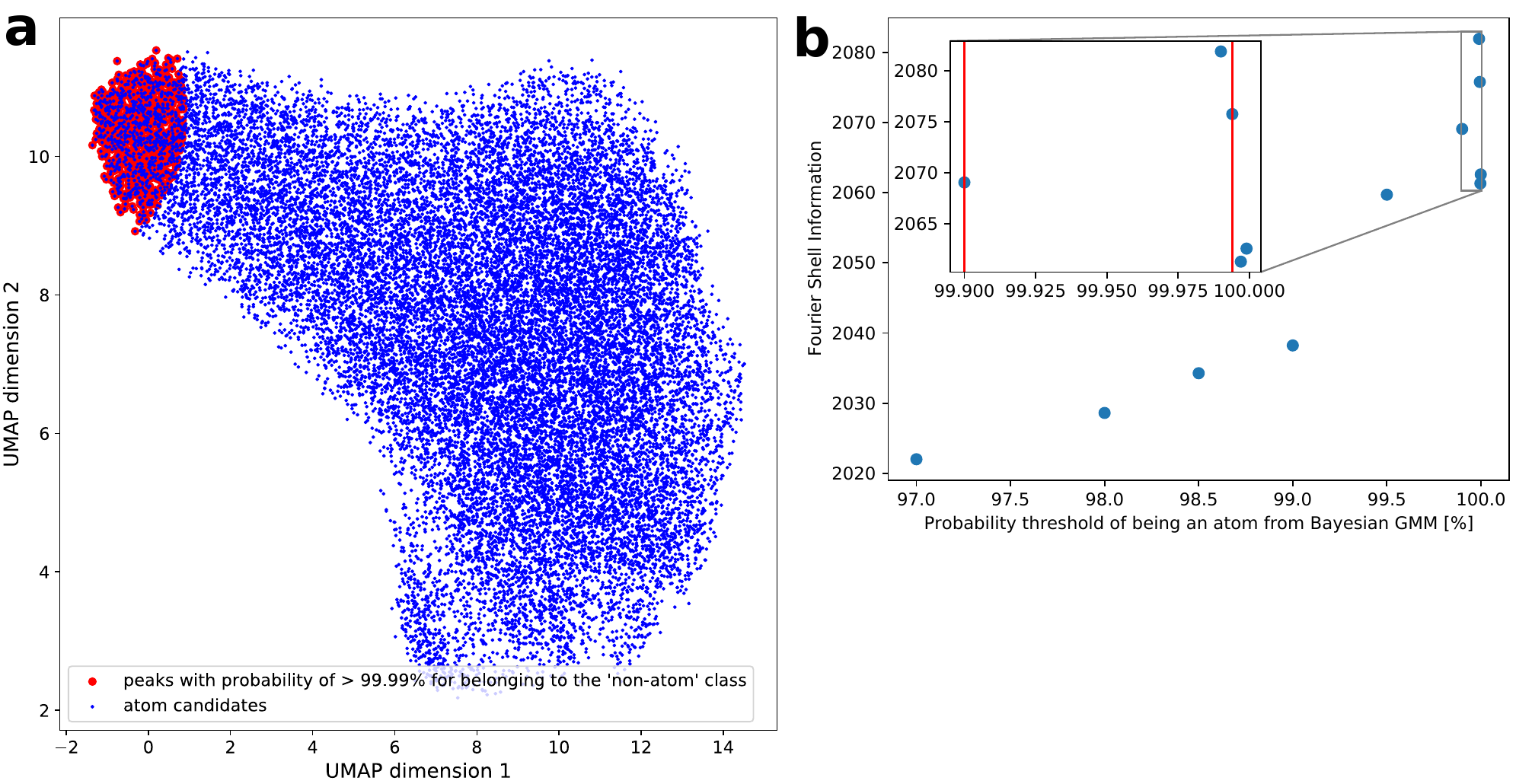}
    \caption{ a) Two-dimensional manifold of the peak candidates obtained with UMAP. b) Fourier shell information of PRISM-simulated tilt series performed with different probability thresholds for the atom candidate list. The probability of and atom candidate not being an atom was obtained from a two-class Bayesian Gaussian Mixture model fitted to the manifold in a). The threshold range between the red lines was identified as the range that extracts the most information from the experimental data.}
    \label{fig:tracing}
\end{figure*}
\newpage
\subsection{STEM Simulations and Tracing Precision}
\begin{figure*}[ht!]
    \includegraphics[width=\textwidth]{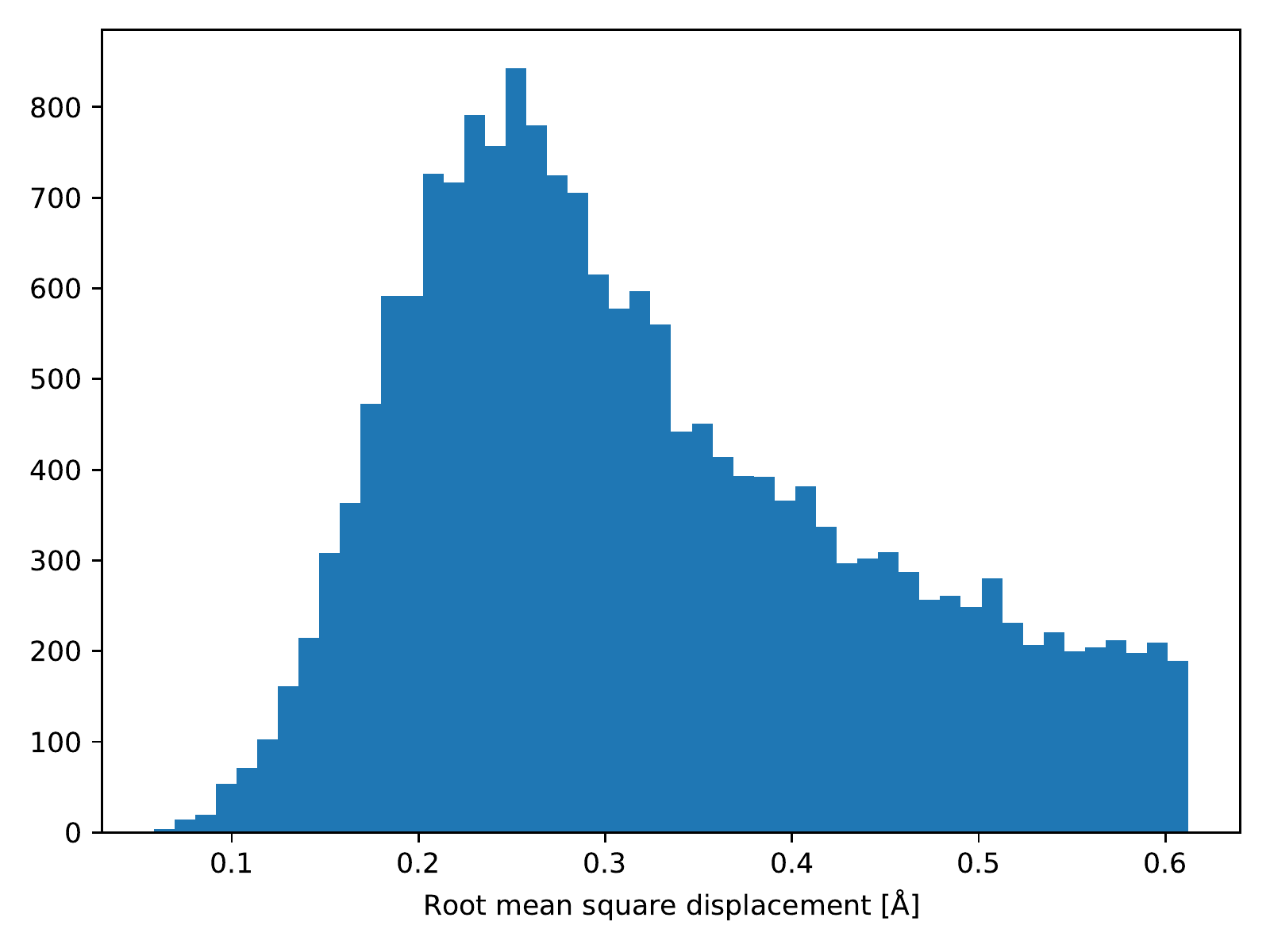}
    \caption{\label{fig:precision} Histogram of the difference in atomic positions between the experimentally determined atomic model and that obtained from multi-slice simulations.}
\end{figure*}

\newpage
\subsection{Ordering Parameter Comparison}
\begin{figure*}[hbt!]
    \centering
    \includegraphics[width=0.6\textwidth]{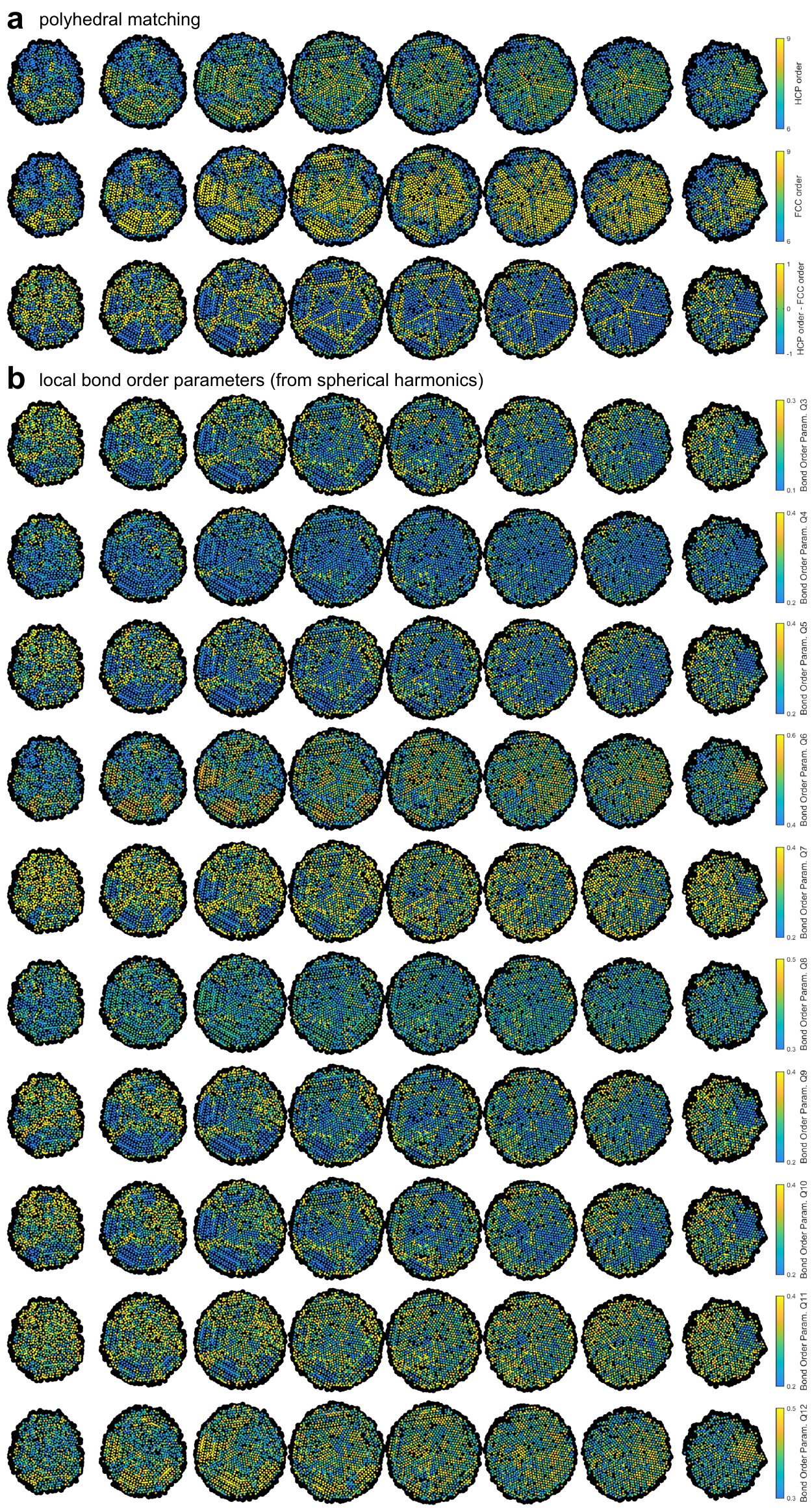}
    \caption{Comparison of methods to determine local ordering. a) Classification of hcp vs fcc ordering in volume slices using a polyhedral matching algorithm \cite{larsen2016robust}. b) Classification of fcc vs hcp ordering in volume slices using spherical harmonics \cite{boo,lechner2008accurate}.}
    \label{fig:ordering_parameter}
 \end{figure*}

% \newpage
% \bibliography{references}
% \end{document}

% \bibliography{references}
\end{document}